\documentclass[preprint,1p]{elsarticle}
\usepackage{lmodern}
\usepackage[T1]{fontenc}
\usepackage{amssymb}
\usepackage{graphics}
\usepackage{amsmath,amsxtra,amssymb,latexsym, amscd,color}
\usepackage[mathscr]{eucal}
\usepackage{array,longtable,tabularx}
\usepackage{multirow}
\usepackage{booktabs} 
\usepackage{pdflscape}
\biboptions{numbers,sort&compress}
\usepackage{hyperref}
\hypersetup{bookmarksnumbered,%
	colorlinks,%
	linkcolor=blue,%
	citecolor=blue,%
	plainpages=false,%
	pdfstartview=FitH}

\journal{Nuclear Physics A}
\begin{document}
\begin{frontmatter}
\title{On the Bohr-Sommerfeld quantization condition and assault frequency in a semiclassical model for $\alpha$ decay}
\author{Le Hoang Chien$^{a,b}$ \footnote{Corresponding author: 
	  {\it lhchien@hcmus.edu.vn}}, Nguyen Tri Toan Phuc$^{a,b}$}
\address{$^a$Department of Nuclear Physics, Faculty of Physics and Engineering Physics, University of Science, Ho Chi Minh City, Vietnam.}
\address{$^b$Vietnam National University, Ho Chi Minh City, Vietnam.}

\begin{abstract}
We study the impacts of the Bohr-Sommerfeld quantization condition and the assault frequency on the $\alpha$ decay half-life within the semiclassical model. The potential between the $\alpha$ particle and daughter nucleus is calculated by the double-folding model using the CDM3Y3 density-dependent nucleon-nucleon interaction with a finite-range exchange term. We show that the proper implementation of the Bohr-Sommerfeld condition leads to a considerable change of the calculated $\alpha$ decay half-life with certain forms of potential. We also propose an alternative treatment for the assault frequency based on the generalized oscillator potential. This description of assault frequency considerably improves the agreement between the calculated $\alpha$ decay half-lives and the experimental data. 
\end{abstract}


\end{frontmatter}
\section{Introduction}

The $\alpha$ radioactivity was first discovered at the very beginning of the nuclear era by Rutherford in 1899 \cite{Rutherford1899}. An arising problem was to explain how the $\alpha$ particle can be emitted spontaneously by some nuclei when its energy is much less than the Coulomb barrier height. In 1928, this microscopic phenomenon, named $\alpha$ decay, was successfully explained by Gamow \cite{Gamow1928} and independently by Gurney and Condon \cite{Gurney1929} using the quantum tunneling mechanism in a semiclassical framework. Based on these pioneering studies, there are a large number of works that have been performed to study the $\alpha$ decay for more than a century \cite{Lovas1998,Pfutzner2012,Belli2019,Qi2019}. In recent years, the $\alpha$ decay is still an actively researched topic that attracted many experimental and theoretical works due to its critical role in the study of superheavy element synthesis \cite{Chowdhury2006,Oganessian2015,Hofmann2016,Cui2018,Santhosh2018,Ismail2020,Singh2021} and the interesting phenomenon of superallowed $\alpha$ decay \cite{Macfarlane1965,Capponi2016,Auranen2018,Clark2020}. Along with the development of experimental facilities, a reliable theoretical model for the $\alpha$ decay is required to predict and extract important physics from the experimental results.

On the theoretical side, many $\alpha$ decay studies have been carried out based on Gamow's semiclassical model involving the quantum tunneling phenomenon \cite{Gamow1928} in combination with the Wentzel-Kramers-Brillouin (WKB) approximation \cite{Buck1992}. In this model, the preformed $\alpha$ particle can penetrate through the Coulomb barrier with a height greater than the total energy of the $\alpha$ particle. The penetration probability is related to the $\alpha$ decay half-life via the concept of the assault frequency (sometimes referred to as the normalization factor) \cite{Kelkar2007}, which is the collision number of the $\alpha$ particle with the Coulomb barrier per second. Therefore, in the study of $\alpha$ decay within the semiclassical framework, one needs to accurately determine the penetration probability and the assault frequency. 

Within the semiclassical model, the penetration probability and the assault frequency are usually estimated by the standard WKB approximation. To properly use the WKB approach in the $\alpha$ decay study, it is necessary to take into account the Bohr-Sommerfeld quantization condition (BSQC) \cite{Buck1992,Kelkar2007}. The implementation of the BSQC ensures the correct behavior of the quasibound wave function of the $\alpha$-daughter system within the semiclassical calculation \cite{Kelkar2007}. This procedure provides a better constraint for the $\alpha$-nucleus potential, which is critical for an accurate $\alpha$ decay description. In addition, the BSQC imposes the Pauli exclusion principle on the $\alpha$ particle by restricting its nucleons to valence orbitals outside the already occupied core \cite{Ni2010-2,Seif2020}. However, the BSQC was not included in many studies of $\alpha$ decay \cite{Sun2016,JDeng2017,Liu2019} which can lead to unreliable estimations of $\alpha$ decay half-life due to the ambiguity of nuclear potential. Thus, in this work, we investigate the critical role of the BSQC in the estimation of $\alpha$ decay half-life and the choice of nuclear potential. 

Generally, the assault frequency in the semiclassical approach is calculated by assuming the $\alpha$ particle inside the parent nucleus bounces between the potential walls. In such a picture, the frequency is inversely proportional to the time required for the $\alpha$ particle to travel back and forth between two walls \cite{Kelkar2007}. Within a quantum mechanical model, the periodic movement of the $\alpha$ particle inside the parent nucleus can be quantized as a quantum harmonic oscillator characterized by a finite angular frequency. Therefore, in this work, we also propose an alternative treatment in which the assault frequency is derived from a quantum harmonic oscillator potential.

In the present work, we perform a theoretical calculation of the $\alpha$ decay half-lives based on the semiclassical framework with the WKB approximation. The nuclear $\alpha$-daughter potential, an important ingredient in the calculation of the $\alpha$ decay half-life, is microscopically constructed within the double-folding model \cite{Khoa2000} using the realistic CDM3Y3 density-dependent nucleon-nucleon (NN) interaction with a finite-range exchange part \cite{Khoa1997}. The goal of this paper is twofold. First, the role of the BSQC and assault frequency from different prescriptions are investigated. Second, we propose an alternative treatment based on the quantum harmonic oscillator picture to calculate the frequency in which the $\alpha$ particle is assumed to move in the harmonic oscillator potential. The outline of the paper is as follows. The details of the BSQC and assault frequency within the semiclassical model for the $\alpha$ decay and the double-folding potential are expressed in Sec. 2. The calculated results and discussions are shown in Sec. 3. Finally, we summarize the conclusions of this work in the last section.
 
\section{Formalism} 

In the framework of the semiclassical model \cite{Gurvitz1987,Buck1992}, the $\alpha$ decay half-life can be determined by the $\alpha$ decay width $\Gamma_\alpha$ as follows
\begin{equation}
T_{1/2} =\frac{\hbar \text{ln} 2}{\Gamma_\alpha},
\label{eq1}
\end{equation}
with 
\begin{equation}
\Gamma_\alpha = \hbar S_\alpha \nu P.
\label{eq2}
\end{equation}
Here $S_\alpha$ is the $\alpha$ preformation factor taken from \cite{JDeng2021}. $\nu$ and $P$ stand for the assault frequency of the $\alpha$ particle at the potential barrier and the penetration probability, respectively. The penetration probability $P$ is evaluated using the standard WKB approximation 
\begin{equation}
 P = \dfrac{1}{1+\exp(x)}~ {\rm with}~x = 2\int_{R_2}^{R_3} \sqrt{\dfrac{2 \mu} {\hbar^2}~
 \Big (V_{\rm T}(R)-Q_{\alpha} \Big )} dR,
\label{eq3}
\end{equation}
where $\mu$ is the reduced mass of the $\alpha$-daughter system. Within the semiclassical model, the assault frequency $\nu$ is defined as the inverse of the time $T$ required for the $\alpha$ particle to move back and forth between the potential walls given as \cite{Kelkar2007,Gurvitz1987,Buck1992}
\begin{equation}
\nu = \frac{1}{T} =\dfrac{\hbar}{2\mu}~\left[ \int_{R_1}^{R_2} \dfrac{dR}{\sqrt{ \dfrac{2\mu}{\hbar^2}\Big ( Q_{\alpha}-V_{\rm T}(R)\Big ) }}
    \right]^{-1}.
\label{eq4}
\end{equation}
Here $R_1$, $R_2$, and $R_3$ in \eqref{eq3} and \eqref{eq4} are the classical turning points obtained from the equations $V_{\rm T}(R_1) = V_{\rm T}(R_2) = V_{\rm T}(R_3) = Q_\alpha$.

To calculate the $\alpha$ decay half-life $T_{1/2}$, one needs to determine the total potential $V_{\rm T}(R)$ at each nuclear distance $R$ consisting of the nuclear, Coulomb potentials, and centrifugal part as  
\begin{equation}
V_{\rm T}(R) = \lambda~V_{\rm N}(R)+ V_{\rm C}(R) + V_{\rm L}(R).
\label{eq5}
\end{equation}
Here the Coulomb potential $V_{\rm C}(R)$ is evaluated by folding two uniform charge distributions \cite{Poling1976} with the chosen charge radii $R^\alpha_{\rm C}=2.2$ fm for the $\alpha$ particle and $R^{\rm d}_{\rm C}=1.2A^{1/3}$ fm for the daughter nuclei that describe reasonably the experimental root-mean-square charge radii \cite{Devries1987}. Such a choice of the Coulomb potential was shown to be accurate up to small internuclear distances \cite{Brandan1997}. The centrifugal potential $V_{\rm L}(R)$ at each orbital angular momentum $L$ is expressed as
\begin{equation}
V_{\rm L}(R) = \dfrac{ \hbar^2 \Big ( L+\dfrac{1}{2} \Big )^2 } {2\mu R^2}.
\label{eq6}
\end{equation}
Here the Langer correction that replaces $L(L+1)$ by $\Big ( L+\dfrac{1}{2} \Big )^2$ is used to obtain the correct behavior of the WKB radial wave function at the origin  \cite{Langer1937}. $L$ values are chosen to satisfy the spin and parity selection rules
\begin{equation}
 | I_i - I_f | \leq  L \leq I_i + I_f {\rm ~~~~ and ~~~~} \dfrac{\pi_i}{\pi_f} = (-1)^L,
\label{eq7}
\end{equation}
where $I_i$ and $I_f$ are the spins of parent and daughter nuclei, respectively, while $\pi_i$ and $\pi_f$ are the corresponding parities. The nuclear potential $V_{\rm N}(R)$ is scaled by a $\lambda$ factor to reproduce the quasibound wave function of $\alpha$-daughter system characterized by the $Q_\alpha$ value. This $\lambda$ factor can be obtained by applying the well-known BSQC \cite{Buck1992,Kelkar2007} that fulfills the periodic motion of the $\alpha$ particle corresponding the quasibound state as follows
\begin{equation}
\int_{R_1}^{R_2} \sqrt{\dfrac{2~\mu}{\hbar^2}~
\Big ( Q_{\alpha} - \lambda~V_{\rm N}(R)- V_{\rm C}(R) - V_{\rm L}(R) \Big )}
dR
= \big (2n+1 \big )\frac{\pi}{2},
\label{eq8}
\end{equation}
where $n$ is the number of internal nodes in the quasibound state of $\alpha$-daughter system that relates to the global quantum number $G$ via the Wildermuth-Tang rule \cite{Wildermuth1977}
\begin{equation}
 G = 2n + L = \sum_{i=1}^4 \big (2 n_i +\ell_i \big ) = \sum_{i=1}^4 g_i.
\label{eq8a}
\end{equation}
The $G$ value in the expression (\ref{eq8a}) is selected to satisfy the Pauli exclusion principle that only allows the nucleons of the $\alpha$ particle with the quantum
numbers $g_i = 2 n_i +\ell_i$ to orbit outside the shell occupied by the nucleons of the daughter nucleus \cite{Wildermuth1977,Ni2010-1}. In other words, the Wildermuth-Tang rule limits the number of internal nodes $n$ to a specific value so that the nucleons composing the $\alpha$ particle occupy the level energies immediately above the Fermi level of the nucleons of the daughter nucleus. Through the ${G}$ values, the shell effect manifestation when the nucleon crosses major shells is also taken into account. We note that the $\lambda$ factor is not a free parameter, but is determined to obtain the quasibound wave function characterized by the $Q_\alpha$ value and satisfied the Pauli exclusion principle.

Because the nuclear part $V_{\rm N}(R)$ in (\ref{eq5}) is not well-known, there are many potential models that have been used for this ingredient. There are only a few studies to derive the $\alpha$-nucleus potential from consistent analyses of elastic scattering and fusion data \cite{Mohr2000,Denisov2005}, and most $\alpha$ decay studies have used the nuclear potentials whose reliability has not been tested on other types of nuclear process \cite{Ghodsi2016,Sun2016,Koyuncu2021}. This can lead to the uncertainty in the calculated $\alpha$ decay half-lives even when constrained with the BSQC. 

In this perspective, one can use a nuclear potential that provides a good description of the elastic scattering data in the optical model analysis to study the $\alpha$ decay. One of such potentials is the well-known double-folding model \cite{Brandan1997,Khoa2000} with a density-dependent NN interaction. The double-folding potential gives a good fit to the elastic angular distribution data for many $\alpha$-nucleus systems, especially for the strongly refractive rainbow pattern that allows investigating the nuclear interaction from the surface to even the interior region \cite{Khoa2007}. It also reproduces reasonably the bound state properties of many $\alpha$-nucleus systems \cite{Mohr2006,Mohr2000}. In addition, this model has been successfully used in many studies of $\alpha$ decay \cite{Mohr2000,Xu2005,Ismail2020}. These results suggest that the double-folding potential with a realistic density-dependent NN interaction is well suited for using in an $\alpha$ decay study. Therefore, in this study, the nuclear potential $V_{\rm N}(R)$ is evaluated by the double-folding model. 

Given an explicitly density- and energy-dependent NN interaction $v(\rho,E,s)$, the direct part of the double-folding potential at a certain distance $R$ can be written in terms of the ground-state densities of the $\alpha$ particle and daughter nucleus, i.e., $\rho_\alpha({\bf r}_\alpha)$ and $ \rho_A({\bf r}_A)$, as follows \cite{Brandan1997,Khoa2000}
\begin{equation}
 V_{\rm D}(E,R)=\int\rho_\alpha({\bf r}_\alpha)\rho_A({\bf r}_A)
 v_{\rm D}(\rho,E,s)d^3r_\alpha d^3r_A, \ \ {\bf s}={\bf r}_A-{\bf r}_\alpha+{\bf R}.
\label{eq10}
\end{equation}
${\bf s}$ stands for the separation between two interacting nucleons having the radii ${\bf r_\alpha}$ and ${\bf r_{\rm A}}$, respectively. $E$ is the center of mass energy.

In general, the exchange term $V_{\rm EX}(E,R)$ is nonlocal, but one can obtain an accurate local form by using the density matrix approximation in \cite{Campi1978} 
\begin{equation}
\begin{split}
V_{\rm EX}(E,R)= \int & \rho_\alpha({\bf r}_\alpha,{\bf r}_\alpha +{\bf s})
 \rho_A({\bf r}_A,{\bf r}_A -{\bf s}) \\ 
 & \times v_{\rm EX}(\rho,E,s)\exp
\left(\frac{i~{\bf K}(E,R)~{\bf s}}{M}\right)d^3r_\alpha d^3r_A. \label{eq11}
\end{split}
\end{equation}
$M=4A/(4+A)$ is the recoil factor with $A$ being
the mass number of daughter nucleus. The local relative-motion momentum ${\bf K}(E,R)$ is determined self-consistently by the following formula
\begin{equation}
 K^2(E,R)={{2\mu}\over{\hbar}^2}[E-V_{\rm D}(E,R)-V_{\rm EX}(E,R)-V_{\rm C}(R)]. \label{eq12}
\end{equation}
In the present work, we use the density-dependent CDM3Y3 interaction \cite{Khoa1997} constructed to include an explicit density dependence into the original M3Y-Paris interaction \cite{Anantaraman1983} as 
\begin{equation}\label{eq13}
 v_{\rm D(EX)}(\rho,E,s)=g(E)F(\rho)v^{\rm D(EX)}_{00}(s). 
\end{equation}  
The radial parts of the direct and exchange terms $v_{00}^{\rm D(EX)}(s)$ were kept 
unchanged as the original M3Y-Paris interaction taken in form \cite{Anantaraman1983}
\begin{eqnarray}\label{eq14} 
 v^{\rm D}_{00}(s)&=&11061.625\dfrac{\exp(-4s)}{4s}-2537.5\dfrac{\exp(-2.5s)}{2.5s} 
 ~{\rm (MeV)}, \\ \nonumber
 v^{\rm EX}_{00}(s)&=&-1524.25\dfrac{\exp(-4s)}{4s}-518.75\dfrac{\exp(-2.5s)}{2.5s} \\ \nonumber
&&-7.8474\dfrac{\exp(-0.7072s)}{0.7072s} 
 ~{\rm (MeV)}. 
\end{eqnarray}  
The density-dependent part $F(\rho)$ is parameterized in terms of the overlapping density $\rho$ as \cite{Khoa1997}
\begin{equation}
F(\rho) = C \Big [  1 + \alpha e^{-\beta \rho}
- \gamma \rho \Big ], \label{eq16}
\end{equation}
where the parameters $C=$ 0.2985, $\alpha=$ 3.4528, $\beta=$ 2.6388 fm$^{3}$, and $\gamma=$ 1.5 fm$^{3}$ were chosen to correctly reproduce the saturation properties of cold symmetric nuclear matter in the Hartree-Fock calculation \cite{Khoa1997}. $g(E) = [1-0.003\varepsilon]$ is the energy-dependent factor \cite{Brandan1997}. The overlapping density $\rho$ in (\ref{eq16}) is taken as the the sum of the ground-state densities of the $\alpha$ particle and daughter nucleus at the position of each interacting nucleon, i.e., $\rho = \rho_\alpha({\bf r}_\alpha)+\rho_A({\bf r}_A)$. 

For simplicity, many calculations use the zero-range approximation for the single-nucleon knock-on exchange, in which the exchange potential $V_{\rm EX}(E,R)$ is included by adding a zero-range pseudo-potential to the interaction $v$ in (\ref{eq10}) \cite{Brandan1997}. In particular, given a density-independent M3Y-Paris interaction $v_{00}(E,s)$ with the zero-range exchange contribution, the double-folding potential is calculated using (\ref{eq10}) with replacing $v_{\rm D}(\rho,E,s)$ by $v_{00}(E,s)$ as
\begin{equation}\label{eq15}
 v_{00}(E,s)= g(E) 
 \Big [ 11061.625\dfrac{\exp(-4s)}{4s}-2537.5\dfrac{\exp(-2.5s)}{2.5s} + \widehat{J}(E)\delta(s) \Big ]
, 
\end{equation}
where $\widehat{J}(E) = -590 \big ( 1-0.002 \varepsilon \big )$ with $\varepsilon = E_{\alpha}/4$ (MeV) is the kinetic energy per nucleon of the $\alpha$ particle.

In the present work, we use the nuclear ground-state density distributions taken from the results of the Hartree-Fock-Bogoliubov method using the BSk14 Skyrme interaction \cite{Goriely2007} to describe the nuclear densities of the daughter nuclei. For the nuclear density distribution of the $\alpha$ particle, the standard Gaussian form $\rho_\alpha (r) = 0.4229\exp(-0.7024r^2)$ \cite{Satchler1979} is chosen. These densities are fold with the density-dependent CDM3Y3 interaction with a finite-range exchange part (\ref{eq14}) \cite{Khoa1997} to obtain the nuclear potential using the equations of (\ref{eq10})-(\ref{eq16}), which we refer to as the CDM3Y3 potential. We also calculate the folding potential with the M3Y-Paris interaction with the zero-range exchange term in (\ref{eq15}) as the M3Y-ZR potential. In the next section, the CDM3Y3 and M3Y-ZR potentials are applied into the semiclassical model for studying the $\alpha$ decay.

\section{Results and discussion}
\subsection{Effects of the BSQC on folding potentials}
We first investigate the effect of the BSQC \eqref{eq8} on the calculated $\alpha$ decay half-life. As seen in \eqref{eq8}, one needs to specify the global quantum number $G$ before applying the semiclassical method to calculate the $\alpha$ decay half-lives. The value of $G$ is chosen so that all nucleons of the $\alpha$ particle occupy at states above those already occupied by nucleons of the daughter nucleus. In the present study, we take the prescription proposed in \cite{Mohr2006,Mohr2007} to determine the $G$ values as
\begin{equation}\label{eq17}
G = \sum_{i=1}^4 {g_{i}},
\end{equation}
with $g_{i} = 4$ for nucleons in the $50 \leq N,Z \leq 82$ shell, $g_{i} = 5$ for nucleons in the $82 < N,Z \leq 126$ shell, and $g_{i} = 6$ for nucleons outside the $N = 126$ shell. This prescription gives the same global quantum number $G$ as the results extracted from the orbital quantum numbers $n_{i}$ and $\ell_{i}$ of $i^{\rm th}$ nucleons from shell model. For the $\alpha+$ $^{208}_{82}$Pb structure in $^{212}_{84}$Po nucleus, there are two protons of the $\alpha$ cluster occupying the $1h_{9/2}$ orbit above the Fermi level of the protons of $^{208}_{82}$Pb daughter nucleus. The $1h_{9/2}$ orbit characterized by $n_i = 0$ and $\ell_i = 5$ has an oscillator quantum number $g_{i} = 5$ with ${i}=$ 1,2. Similarly, two neutron of the $\alpha$ cluster occupy the $2g_{9/2}$ orbit ($n_{i} = 1$ and $\ell_{i} = 4$) corresponding to $g_{i} = 6$ with ${i}=$ 3,4. In such a case, the global quantum number is defined as $G = \sum_{i=1}^4 g_{i} = 22$. 

\begin{table}[h!]
\centering
\caption{The $\alpha$ decay half-lives $T_{1/2}$ obtained from semiclassical model with and without the BSQC. The experimental $\alpha$ decay $Q_\alpha$ values are taken from \cite{MWang2021}.}
\label{tb2} \vspace{0.5cm}
\begin{tabular}{ c  c c c c c c }\toprule
Potential            & $\lambda$ & $\nu$ & $P$ & $P$ \cite{Zhang2011} & $T_{1/2}^{\rm Cal}$ & $T_{1/2}^{\rm Ex}$ \cite{Kondev2021} \\
                     &           &$ ({\rm s}^{-1})$ & & & (s) & (s) \\ \hline 
\multicolumn{7}{c}{$^{210}_{84}$Po $\rightarrow$ $^{206}_{82}$Pb + $\alpha$ 
($Q_{\alpha} =$ 5.408 MeV, $G=20$)} \\ 
\hline
\multirow{2}{4em}{M3Y-ZR}
& 1     & 3.575& 6.15E{--27} &       
\multirow{4}{4.0em}{7.61E{--27}} &    1.047E{+06} & 
\multirow{4}{4.5em}{1.196E{+07}} \\ 
&  0.610& 3.155& 9.99E{--28} &       &    7.305E{+06} & \\
\multirow{2}{4em}{CDM3Y3}
& 1     & 3.045& 8.03E{--27} &       &    9.415E{+05} & \\ 
&  0.849& 2.959& 4.54E{--27} &       &    1.717E{+06} & \\ 
 \hline
\multicolumn{7}{c}{$^{212}_{84}$Po $\rightarrow$ $^{208}_{82}$Pb + $\alpha$ 
($Q_{\alpha} =$ 8.954 MeV, $G=22$)} \\ 
\hline
\multirow{2}{4em}{M3Y-ZR}
& 1     & 3.363& 3.70E{--14} &       
\multirow{4}{4.0em}{4.60E{--14}} &    7.560E{--08} & 
\multirow{4}{4.5em}{2.944E{--07}} \\
&  0.662& 3.098& 9.93E{--15} &       &    3.061E{--07} & \\ 
\multirow{2}{4em}{CDM3Y3}
& 1     & 2.989& 4.80E{--14} &       &    6.567E{--08} & \\ 
&  0.925& 2.855& 3.80E{--14} &       &    8.679E{--08} & \\ 
 \hline
\multicolumn{7}{c}{$^{214}_{84}$Po $\rightarrow$ $^{210}_{82}$Pb + $\alpha$ 
($Q_{\alpha} =$ 7.834 MeV, $G=22$)} \\ 
\hline
\multirow{2}{4em}{M3Y-ZR}
& 1     & 3.473& 4.10E{--17} &      
\multirow{4}{4.0em}{4.31E{--17}} &     5.198E{--05} & 
\multirow{4}{4.5em}{1.635E{--04}} \\ 
&  0.667& 3.184& 1.06E{--17} &      &     2.191E{--04} & \\
\multirow{2}{4em}{CDM3Y3}
& 1     & 2.937& 5.37E{--17} &      &     4.687E{--05} & \\ 
&  0.929& 2.777& 4.27E{--17} &      &     6.251E{--05} & \\ 
 \bottomrule
\end{tabular}
\end{table}

\begin{figure}[htb] \vspace{-1.5cm}
\centering
\includegraphics[width=0.9\textwidth]{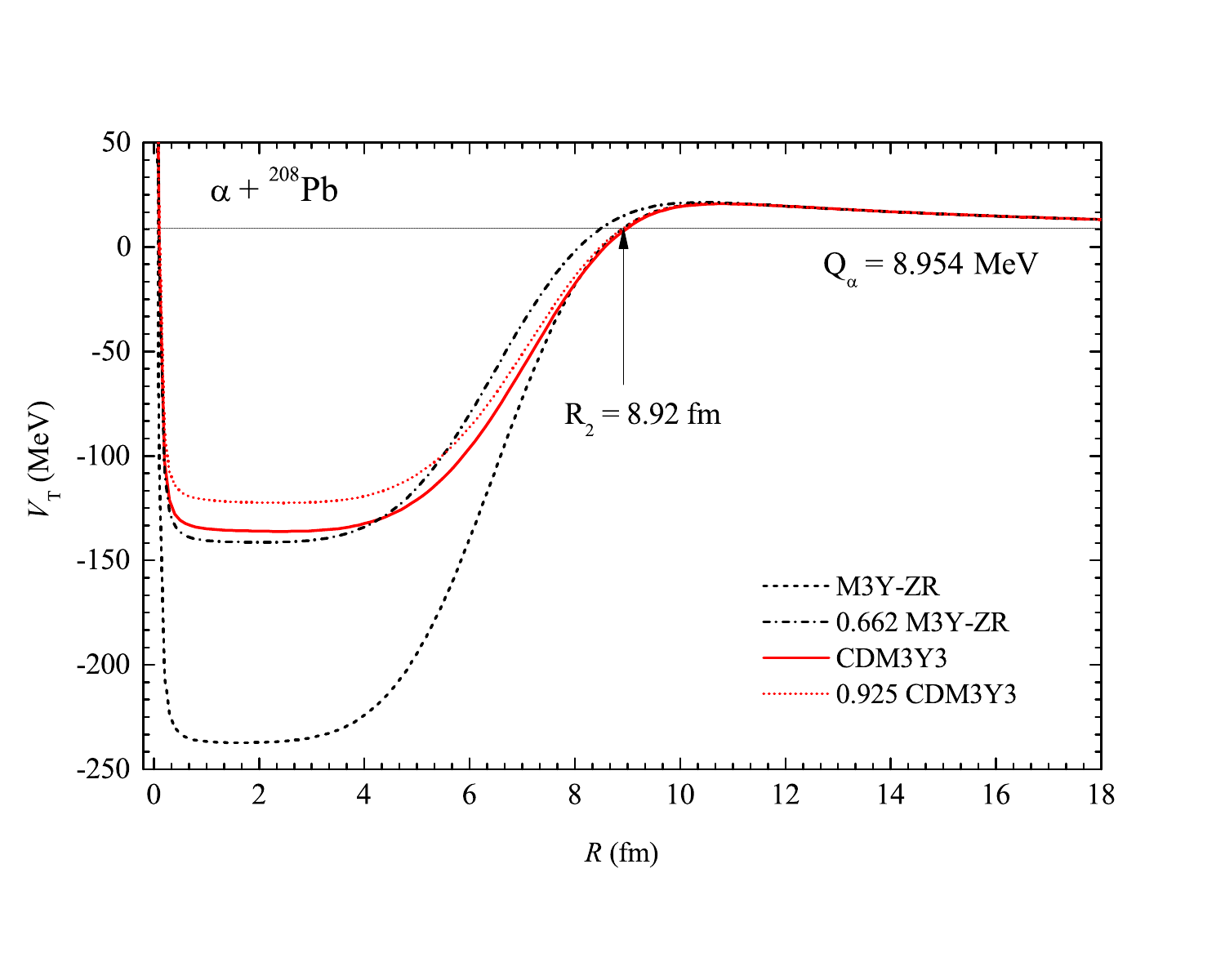} \vspace{-1.2cm} \hspace{-0.5cm}
\caption{The total $\alpha+^{208}$Pb potential at $L=0$ with the nuclear part described by the CDM3Y3 and the M3Y-ZR potential models in two cases of with and without renormalization obeying the BSQC.} \label{f1}
\end{figure}

In Table \ref{tb2}, the calculated results of the penetration probabilities $P$ and $\alpha$ decay half-lives $T_{1/2}$ of the even-even $^{210-214}$Po nuclei are compared with the generalized liquid-drop model (GLDM) results \cite{Zhang2011} and the experimental data \cite{Kondev2021}, respectively. These values are compared with those obtained from the CDM3Y3 and M3Y-ZR double-folding potentials. We choose to compare our results with the GLDM \cite{Zhang2011} due to its capability to reproduce with high accuracy the experimental data for many $\alpha$ emitters. As illustrated in Fig.~\ref{f1}, the M3Y-ZR potential has much deeper strength at the interior region than the one of CDM3Y3 potential due to the lack of density dependence (or medium effect) \cite{Brandan1997}. It is well known that the density dependence of the effective NN interaction is required to reasonably reproduce the cross sections of nucleus-nucleus scattering \cite{Brandan1997,Khoa2007}.
  
Without the implementation of the BSQC ($\lambda = 1$), both potentials provide the penetration probabilities with the same order of magnitude with the results from the GLDM \cite{Zhang2011}. One notes that the penetration probability is mostly sensitive to the Coulomb and the tail of nuclear interactions. As illustrated in Fig.~\ref{f1} for the case of the $\alpha+^{208}$Pb system, the penetration probability is mainly determined by the total interaction at distances larger than the turning point $R_2 = 8.92$ fm, which is almost the same for the CDM3Y3 and the M3Y-ZR potentials. This explains the generally similar results of the penetration probabilities of two potential models with very different strengths at the inner region. With the specified penetration probability, the calculated half-life depends on the assault frequency in (\ref{eq4}) involving the nuclear potential at short distances. However, one can see in Table \ref{tb2} that the frequency is not sensitive to the strength of nuclear potential. Consequently, within the semiclassical model without the implementation of the BSQC, any potential model with the proper shape and strength in the surface region can provide a reasonable description of the experimental $\alpha$ decay half-lives, as seen in Table \ref{tb2}. However, the exact values of these models are still considerably different from each other. The results indicate that applying the semiclassical model without the BSQC can lead to an ambiguity of nuclear potentials used in the study of $\alpha$ decay. We remark that the whole discussion related to the BSQC does not apply to the GLDM since the potentials used in that framework are constrained with different criteria and purposes \cite{Royer2000}.   

Taking into account the BSQC leads to a significant change of the penetration factor and consequently the $\alpha$ decay half-life. In particular, as shown in Table \ref{tb2}, there is a change of the $\alpha$ decay half-lives with a factor of about 1.3$-$1.8 for the CDM3Y3 potential, while this factor is about 4.0$-$6.9 in the case of using the M3Y-ZR potential. This large change is because the BSQC ensures the nuclear potential can support the quasibound nature of the parent nuclei, which mostly happens within the interior region. Thus, it shows that the BSQC should not be neglected in the consistently semiclassical models for the study of $\alpha$ decay. To obtain the reliable $\alpha$ decay half-life results based on the semiclassical model with the implementation of the BSQC, the used nuclear potential is required to simultaneously satisfy two conditions of providing a reliable tunneling probability and reproducing precisely the oscillation behavior of the quasibound wave function of the $\alpha$-daughter system. This means that the potential should have the proper shape and strength over a wide range of distance from the surface to the interior region. 

In Table \ref{tb2}, the M3Y-ZR potential with very deep strength is renormalized by a factor $\lambda \simeq 0.610-0.667$ to correctly reproduce the number of internal nodes in the quasibound wave function of the $\alpha$-daughter system obeying the BSQC. The renormalization results in a wider Coulomb barrier than the one in the case of neglecting the BSQC, as illustrated in Fig.~{\ref{f1}}. Consequently, this leads to the decrease of penetration probabilities by a factor of about 4 that are significantly different from the GLDM results \cite{Zhang2011}. Therefore, applying the folding M3Y-ZR potential with the renormalized factor $\lambda$ significantly smaller than the unity can lead to an unrealistic result of $\alpha$ decay half-life. 

For the CDM3Y3 potential with the more proper strength and shape in both the surface and interior region in comparison with the M3Y-ZR potential, it requires a lightly renormalized factor of about $0.849-0.929$ to satisfy the BSQC. With a slight change in the strength and shape, the CDM3Y3 potential still gives the proper penetration probabilities and consequently provides the $\alpha$ decay half-lives agreeable to the experimental data, as shown in Table \ref{tb2}. These results indicate that the study of $\alpha$ decay within the semiclassical model taken into account the BSQC can provide a stringent test for the $\alpha$-daughter potential model. Within the semiclassical model for studying the $\alpha$ decay, to obtain realistic half-lives one should use the nuclear potential with the proper strength and shape over a wide range of interaction to provide the reasonable penetration probability as well as satisfy the BSQC with the renormalized factor $\lambda$ close to the unity. For the folding-type potential, the more realistic density-dependent NN interaction should be preferred over the bare M3Y one.

\subsection{Treatments of the assault frequency}
Along with the realistic penetration probability, the assault frequency is one of the most important ingredients to obtain a reliable evaluation of the $\alpha$ decay half-lives. In the literature, the assault frequency is often estimated using the semiclassical model that is inversely proportional to the time period for the $\alpha$ particle to traverse back and forth along with the nuclear distance between the turning points $R_1$ and $R_2$ as expressed in (\ref{eq4}). Here we name such frequency as $\nu^{\rm SC}$ and is written as follows \cite{Kelkar2007,Gurvitz1987,Buck1992} 
\begin{equation}
\nu^{\rm SC} = \frac{1}{T} =\dfrac{\hbar}{2\mu}~\bigg [ \int_{R_1}^{R_2} \dfrac{1}{\sqrt{ \dfrac{2\mu}{\hbar^2}\Big ( Q_{\alpha}-V_{\rm T}(R)\Big ) }}~dR
    \bigg ]^{-1}.
\label{eq18}
\end{equation} 
The assault frequency can be estimated from a different viewpoint in quantum mechanics \cite{Dong2010,Zhang2011} by using the global quantum number and the empirical nuclear radius. In this picture, the $\alpha$ particle is assumed to vibrate near the surface of the parent nucleus in a harmonic oscillator potential, the assault frequency is given in the form \cite{Dong2010,Zhang2011}
\begin{equation}
\nu^{\rm HO} = \dfrac{\Big ( G + \dfrac{3}{2} \Big ) \hbar}
{1.2 \pi \mu R_{\rm N}^2}.
\label{eq20}
\end{equation} 
Here the nuclear radius $R_{\rm N}$ is defined as
\begin{equation}
R_{\rm N} = 1.240 A^{1/3} \Big (
1.0 + \dfrac{1.646}{A} - 0.191 \dfrac{A_{\rm p}-2Z_{\rm p}}{A_{\rm p}}
\Big ),
\label{eq21}
\end{equation} 
where $A_{\rm p}$ and $Z_{\rm p}$ are the mass and atomic number of parent nucleus, respectively. We note that Eqs.~(\ref{eq20}), (\ref{eq21}) prescription of the assault frequency is widely used in the GLDM calculations. 

\begin{table}[bht]
\centering
\caption{The assault frequency obtained from different models of $\nu^{\rm SC}$ (\ref{eq18}), $\nu^{\rm HO}$ (\ref{eq20}), and $\nu^{\rm GOP}$ (\ref{eq24}). The frequency is multiplied by a factor of 10$^{21}$.}
\label{tb3} \vspace{0.5cm}
\begin{tabular*}{\textwidth}{@{\extracolsep{\fill}} c @{\extracolsep{\fill}} c @{\extracolsep{\fill}} c @{\extracolsep{\fill}} c @{\extracolsep{\fill}} c @{\extracolsep{\fill}} c @{\extracolsep{\fill}}} 
\toprule
$\alpha$ process & ${G}$ & $Q_\alpha$ \cite{MWang2021} & {$\nu^{\rm SC}$} 
 & {$\nu^{\rm HO}$}  & {$\nu^{\rm GOP}$}   \\ \cline{4-6}
 &  &  (MeV) & \multicolumn{3}{c}{$(s^{-1})$}    \\ \midrule
{$^{210}_{84}$Po} $\rightarrow$ {$^{206}_{82}$Pb} &20 & 5.408 &
 2.959 & 1.809 & 1.038  \\ 
{$^{212}_{84}$Po} $\rightarrow$ {$^{208}_{82}$Pb} &22 & 8.954 &
 2.855 & 1.970 & 0.949   \\ 
{$^{214}_{84}$Po} $\rightarrow$ {$^{210}_{82}$Pb} &22 & 7.834 &
 2.777 & 1.963 & 1.003   \\ \bottomrule 
\end{tabular*}
\end{table}

In the present work, following a similar approach of \cite{Dong2010}, we attempt to evaluate the assault frequency using the total potential $V_{\rm T} (R)$ in \eqref{eq5}, instead of using the empirical values as \eqref{eq21}. We first remind that the classically confined $\alpha$ particle would move back and forth inside the well with a finite frequency. We use the idea that the $\alpha$ particle oscillates at the vicinity of the stable equilibrium point in the harmonic oscillator potential before tunneling through the Coulomb barrier. For such a case, one can approximate the total potential $V_{\rm T} (R)$ around the equilibrium $R_{\rm e}$ as a harmonic oscillator potential by the Taylor expansion as follows
 \begin{equation}
V_{\rm T}(R) = V_{\rm T}(R_{\rm e}) + 
(R-R_{\rm e}) \dfrac{{\rm d} V_{\rm T}(R)}{{\rm d} R} \bigg |_{R = R_{\rm e}} +
\dfrac{1}{2}(R-R_{\rm e})^2 \dfrac{{\rm d^2} V_{\rm T}(R)}{{\rm d^2} R}
\bigg |_{R = R_{\rm e}} .
\label{eq22}
\end{equation}
Here the equilibrium $R_{\rm e}$ is the position where the total potential $V_{\rm T}(R)$ in \eqref{eq5} is minimum that leads the second term of \eqref{eq22} to be vanished. The last term is defined as the harmonic oscillator potential that can be written in terms of the angular frequency $\omega$ as \cite{Landau1977}
 \begin{equation}
 \begin{split}
\dfrac{1}{2}(R-R_{\rm e})^2 \dfrac{{\rm d^2} V_{\rm T}(R)}{{\rm d^2} R} \bigg |_{R = R_{\rm e}}
= \dfrac{1}{2} (R-R_{\rm e})^2 \mu \omega^2  
.
\end{split}
\label{eq23}
\end{equation}
The relation in \eqref{eq23} can be used to determine the generalized oscillator potential (GOP) assault frequency as follows
\begin{equation}
\nu^{\rm GOP} = \dfrac{\omega}{2\pi}
=\dfrac{1}{2 \pi}~ \Bigg [
\dfrac{1}{\mu} \dfrac{{\rm d^2} V_{\rm T}(R)}{{\rm d^2} R} \bigg |_{R = R_{\rm e}}  
\Bigg ]^{1/2}.
\label{eq24}
\end{equation}

We compare the assault frequencies from different prescriptions in Table \ref{tb3}. The frequency $\nu^{\rm SC}$ stands for the result of the semiclassical model using (\ref{eq18}). $\nu^{\rm HO}$ is the harmonic oscillator-based frequency from (\ref{eq20}) using the global quantum number $G$ given in Table \ref{tb3} while $\nu^{\rm GOP}$ is the frequency proposed in the present work using (\ref{eq24}). Throughout this calculation, we use the CDM3Y3 potential to describe the nuclear part $V_{\rm N}(R)$ of the total potential $V_{\rm T}(R)$ in (\ref{eq5}) that is renormalized by a factor $\lambda$ to satisfy the BSQC before used to evaluate the assault frequency. As shown in Table \ref{tb3}, the order of magnitude of the assault frequency $\nu^{\rm GOP}$ calculated from the generalized oscillator potential prescription agrees reasonably with these of the $\nu^{\rm SC}$ and $\nu^{\rm HO}$ values obtained from the previous models. Both $\nu^{\rm HO}$ and $\nu^{\rm GOP}$ results based on the quantum oscillator picture are smaller than the $\nu^{\rm SC}$ values obtained from the semiclassical model. The $\nu^{\rm GOP}$ values are about half of the $\nu^{\rm HO}$ values. It is difficult to prove which model is better but the $\nu^{\rm GOP}$ prescription proposed in this work can be a reasonable alternative for describing the assault frequency in the semiclassical model for the $\alpha$ decay study.

\begin{figure}[thb] \vspace{-1.0cm} 
\centering
\includegraphics[width=0.92\textwidth]{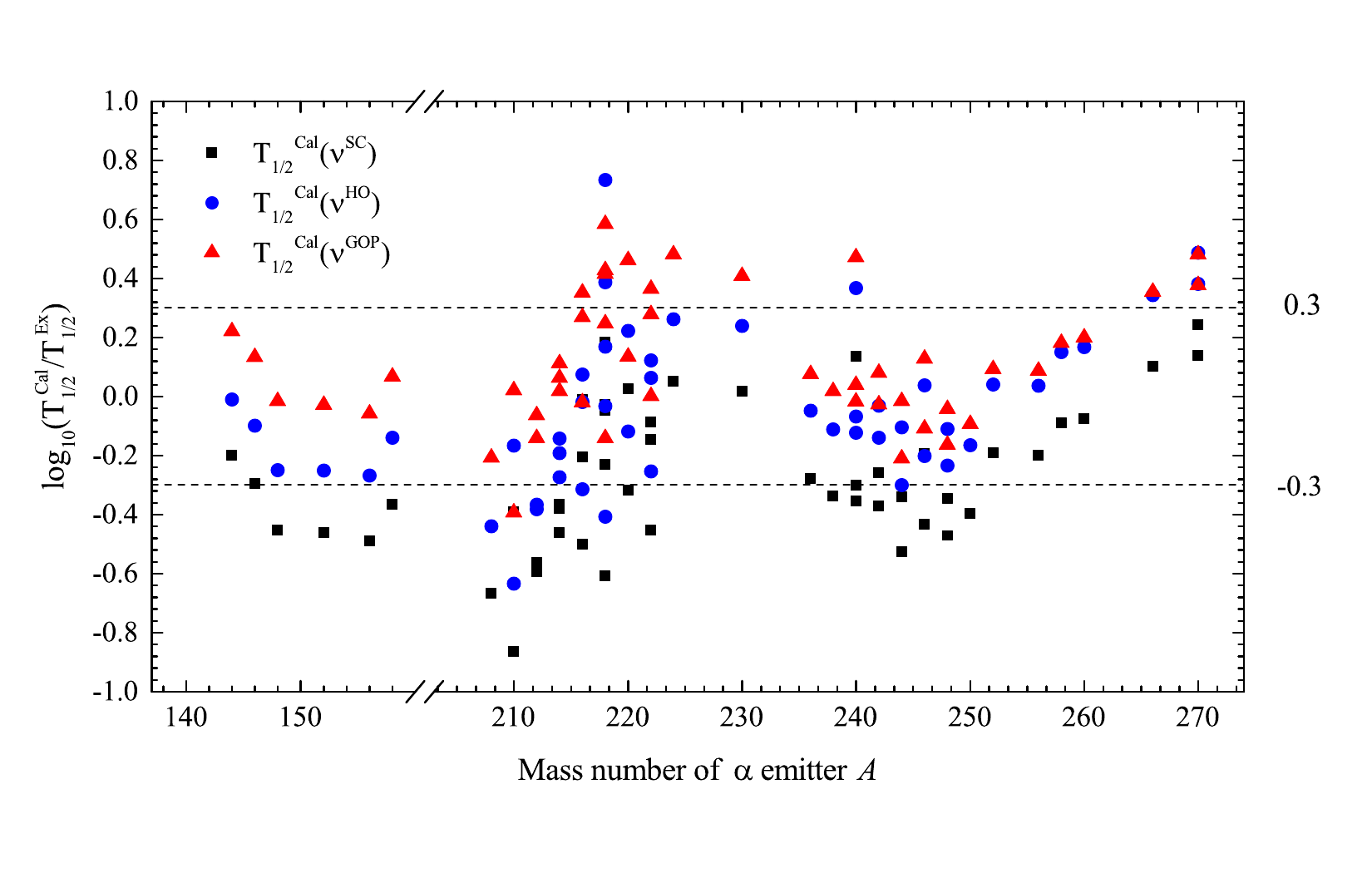} \vspace{-1.0cm} \hspace{-0.5cm}
\caption{The deviations between the decimal logarithms of experimental half-lives and
the calculated results obtained from the semiclassical model using sequentially the $\nu^{\rm SC}$ (\ref{eq18}), $\nu^{\rm HO}$ (\ref{eq20}), and $\nu^{\rm GOP}$  (\ref{eq24}) models.} \label{f2}
\end{figure}

To investigate the validity of our new assault frequency prescription $\nu^{\rm GOP}$ derived from the generalized oscillator potential, we calculate the $\alpha$ decay half-life within the semiclassical model using the assault frequencies $\nu^{\rm SC}$ (\ref{eq18}), $\nu^{\rm HO}$ (\ref{eq20}), and $\nu^{\rm GOP}$ (\ref{eq24}). The calculated results are compared with the experimental $\alpha$ decay half-lives \cite{Kondev2021} of 50 even-even $\alpha$ emitters with the masses varying from 144 to 270, with the values listed in Table \ref{tb4} in the appendix. To reduce the uncertainty coming from the lack of coupled-channels effect in our model, we only select near spherical nuclei with quadrupole deformation close to zero. 

The deviations of the decimal logarithm of $\alpha$ decay half-lives between the calculated results and the experimental data via the mass number of the parent nuclei have been plotted in Fig.~\ref{f2} with $T^{\rm Cal}_{1/2}({\nu^{\rm GOP}})$, $T^{\rm Cal}_{1/2}({\nu^{\rm SC}})$, and $T^{\rm Cal}_{1/2}({\nu^{\rm HO}})$ standing for the calculated $\alpha$ decay half-life using the assault frequency $\nu^{\rm GOP}$ (\ref{eq24}), $\nu^{\rm SC}$ (\ref{eq18}), and $\nu^{\rm HO}$ (\ref{eq20}), respectively. 

To present a better comparison between the three options of assault frequency, we employ the commonly used root-mean-square error (RMSE), also know as the root-mean-square deviation, of the half-life decimal logarithm \cite{Poe06,Denisov2009}
\begin{equation}
 {\rm RMSE} = \sqrt{
 \dfrac{1}{N-1} \sum_{i=1}^N
\Bigg [ 
 {\rm \log_{10}}~ T_{1/2,i}^{\rm Cal} - {\rm \log_{10}}~ T_{1/2,i}^{\rm Ex}  \Bigg ]^2},  
 \label{eqrmse}
\end{equation}
where $N=50$ is the number of the considered $\alpha$ emitters in Table \ref{tb4}. The RMSE is widely used in systematic $\alpha$-decay studies due to its compatibility with the $\chi^2$ fitting technique. Since the experimental half-life values vary widely across the considered nuclei, we propose that the deviation between theoretical and experimental values can be illustrated in a more meaningful way by using the root-mean-square relative error (RMSRE) given by 

\begin{equation}
	{\rm RMSRE} = \sqrt{
		\dfrac{1}{N-1} \sum_{i=1}^N
		\Bigg [ 
		\dfrac{{\rm \log_{10}}~ T_{1/2,i}^{\rm Cal} - {\rm \log_{10}}~ T_{1/2,i}^{\rm Ex}}
		{{\rm \log_{10}}~ T_{1/2,i}^{\rm Ex}}
		\Bigg ]^2
	}.  
	\label{eqrmsre}
\end{equation}
We use both the RMSE and RMSRE in our discussion below.

As shown in Fig.~\ref{f2}, all three calculated results are in generally good agreement with the experimental data. Two horizontal lines at $\pm$0.3 in Fig.~\ref{f2} that correspond to the deviations of the $\alpha$ decay half-lives within a factor of about 2 are used to guide eyes in the comparison of three assault frequency prescriptions. The ${\rm log}_{10}({T_{1/2}^{\rm Cal}}/{T_{1/2}^{\rm Ex}})$ values for all three cases vary in the range from $-0.8$ to $0.8$ corresponding to the deviations of the $\alpha$ decay half-lives within a factor of about 6.3. This means that the semiclassical model using the ${\nu^{\rm SC}}$, ${\nu^{\rm HO}}$, and ${\nu^{\rm GOP}}$ frequencies provides a reasonable description of the experimental data. The RMSE corresponding to the ${\nu^{\rm SC}}$, ${\nu^{\rm HO}}$, and ${\nu^{\rm GOP}}$ are 0.343, 0.248, and 0.259, respectively. These RMSE values are comparable or slightly better than those reported in other systematic studies for even-even chains such as 0.309 \cite{Denisov2009} and 0.516 \cite{Das2007}. Although this comparison is not completely fair due to the smaller number of nuclei considered in our present work, it nevertheless demonstrates the validity of our $\alpha$-decay calculation using the BSQC-applied CDM3Y3 folding potential. 

One can see that the ${\rm log}_{10}({T_{1/2}^{\rm Cal}}/{T_{1/2}^{\rm Ex}})$ values in both cases of using $T^{\rm Cal}_{1/2}({\nu^{\rm HO}})$ and $T^{\rm Cal}_{1/2}({\nu^{\rm GOP}})$ are mostly within the range of about $\pm$0.3 and are closer to the zero than these obtained from the case of using ${T^{\rm Cal}_{1/2}({\rm \nu^{SC}})}$. Indeed, the ${\nu^{\rm HO}}$ and ${\nu^{\rm GOP}}$ frequencies respectively yield similar RMSE values in (\ref{eqrmse}) of 0.248 and 0.259, while the ${\nu^{\rm SC}}$ frequency gives a larger value of 0.343. The situation is the same when using the RMSRE with values corresponding to ${\nu^{\rm SC}}$, ${\nu^{\rm HO}}$, and ${\nu^{\rm GOP}}$ are 0.201,	0.128, and 0.122, respectively. It indicates that the semiclassical preformed cluster model using the assault frequencies of ${\nu^{\rm HO}}$ and ${\nu^{\rm GOP}}$ derived from the harmonic oscillator picture can improve considerably the agreement between the theoretical $\alpha$ decay half-lives and the experimental values in comparison with the case using the ${\nu^{\rm SC}}$ frequency. 

As shown in Fig.~\ref{f2}, the deviation of the $\alpha$ decay half-lives using the ${\nu^{\rm GOP}}$ frequency is comparable with that using the ${\nu^{\rm HO}}$ frequency. This similar deviation is clearly reflected by the close RMSE and RMSRE values of these two descriptions. It is interesting to note that, in contrast to the RMSE case, the RMSRE value of the calculations using our proposed ${\nu^{\rm GOP}}$ frequency is slightly better than that with ${\nu^{\rm HO}}$ definition. Overall, these results imply that the prescription of ${\nu^{\rm GOP}}$ in (\ref{eq24}) derived from the generalized oscillator potential is reliable enough to describe the assault frequency in the semiclassical model for the $\alpha$ decay study.

\begin{table}[thb]
\caption{The $\alpha$ preformation factor $S_\alpha$ is evaluated by (\ref{eq25}) using  different frequency models of $\nu^{\rm SC}$ \eqref{eq18}, $\nu^{\rm HO}$ \eqref{eq20}, and $\nu^{\rm GOP}$ \eqref{eq24}.}
\label{tb5} \vspace{0.5cm}
\begin{tabular*}{\textwidth}{@{\extracolsep{\fill}} c @{\extracolsep{\fill}} c @{\extracolsep{\fill}} c @{\extracolsep{\fill}} c @{\extracolsep{\fill}} c @{\extracolsep{\fill}} c @{\extracolsep{\fill}} c @{\extracolsep{\fill}}c @{\extracolsep{\fill}}} 
\toprule
$\alpha$ & ${G}$ &$\lambda$ & $Q_\alpha$ \cite{MWang2021} & \multicolumn{3}{c}{$S_\alpha$} &  $S_\alpha$ \cite{Yang2020} 
\\ \cline{5-7}
emitter  &    &    & (MeV) & {$\nu^{\rm SC}$} 
 & {$\nu^{\rm HO}$}  & {$\nu^{\rm GOP}$} & 
 \\ \midrule
{$^{104}_{52}$Te} &16 & 0.844 & 5.100 &
 0.455 & 0.634 & 0.836  & 0.723\\ 
{$^{106}_{52}$Te} &16 & 0.840 & 4.290 &
 0.186 & 0.254 & 0.333  & \\ 
{$^{210}_{84}$Po} &20 & 0.849 & 5.408 &
 0.004 & 0.007 & 0.012  & 0.013\\ 
{$^{212}_{84}$Po} &22 & 0.925 & 8.954 &
 0.021 & 0.031 & 0.065  & 0.105\\  \bottomrule
\end{tabular*}
\end{table}
 
The phenomenon of superallowed $\alpha$ decay is an interesting topic that attracts many experimental and theoretical works \cite{Macfarlane1965,Capponi2016,Auranen2018,Clark2020}. This phenomenon implies that there is a significant enhancement of $\alpha$ decay in the region around the doubly magic ($N=Z$) $^{100}$Sn nucleus. In fact, the recent experimental study \cite{Auranen2018} shows that $^{104}$Te nucleus decaying towards $^{100}$Sn nucleus has a significantly large $\alpha$ preformation factor when compared with $^{212}$Po nucleus. Therefore, it is helpful to apply the semiclassical model using the assault frequency $\nu^{\rm GOP}$ derived from the generalized oscillator potential to predict the $\alpha$ preformation factor $S_\alpha$ of $^{104}$Te ($\alpha+^{100}$Sn) and $^{212}$Po ($\alpha+^{208}$Pb) nuclei, and their neighboring $^{106}$Te ($\alpha+^{102}$Sn) and $^{210}$Po ($\alpha+^{206}$Pb) nuclei. 

For this specific part, the $\alpha$ preformation factor $S_\alpha$ is extracted by the ratio of the calculated $T_{1/2}^{\rm Cal}$ value to the experimental $T_{1/2}^{\rm Ex}$ one. Namely,
\begin{equation}\label{eq25}
S_\alpha = \dfrac{T_{1/2}^{\rm Cal}}{T_{1/2}^{\rm Ex}} \quad {\rm with} \quad T_{1/2}^{\rm Cal} =\dfrac{\rm ln2}{\nu P}. 
\end{equation}
$P$ is calculated by using (\ref{eq3}) while $\nu$ is estimated by using $\nu^{\rm SC}$ in (\ref{eq18}), $\nu^{\rm HO}$ in (\ref{eq20}), and $\nu^{\rm GOP}$ in (\ref{eq24}). The experimental $\alpha$ decay half-lives of $^{104}$Te \cite{Auranen2018}, $^{106}$Te \cite{Capponi2016}, $^{210}$Po, and $^{212}$Po \cite{Kondev2021} are then inserted into (\ref{eq25}) to extract the values of $S_\alpha$, listed in Table \ref{tb5}. One can see that the $S_\alpha$ results from using the $\nu^{\rm GOP}$ frequency is generally closer to the predictions of the microscopic calculation \cite{Yang2020} than these from using the $\nu^{\rm SC}$ and $\nu^{\rm HO}$ frequencies.

For $^{104}$Te nucleus, as seen in Table \ref{tb5}, the $\alpha$ preformation factor obtained from (\ref{eq25}) using the $\nu^{\rm GOP}$ frequency is significantly enhanced in comparison with that of $^{212}$Po nucleus. 
The calculated result is consistent with the observed phenomenon of superallowed $\alpha$ decay in $^{104}$Te \cite{Auranen2018} that was believed to have arisen by the enhanced interactions between the valence neutrons and protons in the identical single-particle orbitals \cite{Macfarlane1965,Auranen2018,Clark2020}.
In analogy to $^{104}$Te nucleus, $^{212}$Po undergoes the $\alpha$ decay to doubly magic $^{208}$Pb 
but it has the valence nucleons occupying different orbitals that lead to a smaller $\alpha$ preformation factor.
The extracted $S_\alpha$ values of the neighboring $^{106}$Te and $^{210}$Po nuclei 
using $\nu^{\rm GOP}$ are smaller than these of $^{104}$Te and $^{212}$Po, respectively. The results agree with the expectation of the enhanced $\alpha$ preformation factor in the decay to the doubly magic nuclei. 
These obtained results show that the $\alpha$ preformation factors extracted from (\ref{eq25}) using the $\nu^{\rm GOP}$ frequency vary with the same trend with these calculated by the microscopic method \cite{Yang2020} and are agreement with the discussions of previous studies \cite{Macfarlane1965,Capponi2016,Auranen2018,Clark2020}. This once again implies that the $\nu^{\rm GOP}$ frequency evaluated with the generalized oscillator potential approach is reliable and can be used as an alternative method to describe the assault frequency in the semiclassical model for the $\alpha$ decay study.

\section{Summary}
In this work, we investigate the effects of the BSQC and assault frequency within the semiclassical model on the $\alpha$ decay half-lives. The $\alpha$-daughter potentials are constructed within the framework of the double-folding model using the realistic density- and energy-dependent CDM3Y3 interaction with a finite-range exchange term. 

In the semiclassical model without the BSQC, it is shown that any nuclear $\alpha$-daughter potential model with the proper shape and strength in the surface region can achieve a certain degree of agreement with the experimental $\alpha$ decay half-lives, leading to an ambiguity of the nuclear potentials used in the study of $\alpha$ decay. Taking into account the BSQC results in a significant change of the $\alpha$ decay half-lives that indicates that one should not neglect the BSQC in a reliable semiclassical calculation. To satisfy the BSQC, the strength of the MY3-ZR potential calculated by the folding model using the density-independent M3Y-Paris interaction with the zero-range exchange term must be scaled with a significantly small factor of about 0.610$-$0.667 that can cause the unreasonable result of penetration probability, and consequently the unrealistic $\alpha$ decay half-life. Based on the obtained results, it is shown that the semiclassical model compatible with the BSQC can provide a stringent test for a nuclear $\alpha$-daughter potential over a wide interacting range.

We have proposed an alternative treatment for the assault frequency, which we call the generalized oscillator potential approach. In terms of the quantum oscillator picture without using the empirical values, the alternative treatment applied in the semiclassical model significantly improves the agreement between the theoretical $\alpha$ decay half-lives and the experimental data in comparison with the conventional semiclassical frequency. Moreover, the $\alpha$ preformation factors calculated with this assault frequency for some transitions to doubly magic nuclei are consistent with results from the microscopic study. The present results demonstrate that the generalized oscillator potential description of the assault frequency gives reliable results and can be applied to study the $\alpha$ decay process.

\section*{Acknowledgements}
We thank Prof. Dao T. Khoa for useful discussion related to the double-folding model. The present research was supported, in part, by Vietnam National University, Ho Chi Minh City, Vietnam under the grant C2019-18-02. 

\appendix
\section{$\alpha$-decay half-lives with different assault frequencies}

The frequency $\nu^{\rm SC}$ stands for the result of the semiclassical model using (\ref{eq18}). $\nu^{\rm HO}$ is the quantum oscillator-based frequency from (\ref{eq20}) using the global quantum number $G$ given in Table \ref{tb3} and $\nu^{\rm GOP}$ is the frequency proposed in the present work using (\ref{eq24}).

\begin{longtable}{ c c c c c c c c } 
	\caption{The $\alpha$ decay half-lives calculated within the semi-classical model using $\nu^{\rm SC}$ (\ref{eq18}), $\nu^{\rm HO}$ (\ref{eq20}), and $\nu^{\rm GOP}$ (\ref{eq24}) are compared with the experimental data \cite{Kondev2021}.} 
	\label{tb4} \\ \toprule
	$\alpha$ & $Q_{\alpha}$ \cite{MWang2021} & $T_{1/2}^{\rm Ex}$ \cite{Kondev2021} &  $G$ &  $\lambda$  
	& $T_{1/2}({\nu^{\rm SC}})$ & $T_{1/2}({\nu^{\rm HO}})$  
	& $T_{1/2}({\nu^{\rm GOP}})$  \\
	emitter & (MeV)        & $(s)$                &     &          
	& $(s)$      & $(s)$                           & $(s)$  \\  \midrule 
	\endfirsthead
	\multicolumn{8}{c}%
	{\tablename\ \thetable\ -- \textit{Continued}} \\
	\hline
	$\alpha$ & $Q_{\alpha}$ \cite{MWang2021} & $T_{1/2}^{\rm Ex}$ \cite{Kondev2021} &  $G$ &  $\lambda$  
	& $T_{1/2}({\nu^{\rm SC}})$ & $T_{1/2}({\nu^{\rm HO}})$  
	& $T_{1/2}({\nu^{\rm GOP}})$  \\
	emitter & (MeV)        & $(s)$                &     &          
	& $(s)$      & $(s)$                           & $(s)$  \\  \midrule 
	\endhead
	\multicolumn{8}{r}{\textit{Continued on next page}} \\
	\endfoot
	\endlastfoot
	$^{144}_{60}$Nd& 1.901 & 7.227E+{22} & 18 & 0.866    
	& 4.651E+{22} & 7.145E+{22}  
	& 1.218E+{23} \\ 
	$^{146}_{62}$Sm& 2.529 & 2.146E+{15} & 18 & 0.860    
	& 1.097E+{15} & 1.725E+{15}  
	& 2.950E+{15} \\ 
	$^{148}_{64}$Gd& 3.271 & 2.237E+{09} & 18 & 0.855    
	& 8.295E+{08} & 1.275E+{09}  
	& 2.186E+{09} \\ 
	$^{152}_{68}$Er& 4.934 & 1.139E+{01} & 18 & 0.843    
	& 4.233E+{00} & 6.461E+{00}  
	& 1.080E+{01} \\ 
	$^{156}_{72}$Hf& 6.026 & 2.364E{--02} & 18 & 0.838
	& 9.517E{--03} & 1.298E{--02}  
	& 2.098E{--02}  \\ 
	$^{158}_{74}$W & 6.613 & 1.430E{--03} & 18 & 0.835
	& 6.125E{--04} & 9.184E{--04}  
	& 1.480E{--03} \\ 
	$^{208}_{84}$Po& 5.215 & 9.145E+{07}  & 20 & 0.856
	& 2.092E+{07} & 3.360E+{07}  
	& 5.752E+{07} \\ 
	$^{210}_{84}$Po& 5.408 & 1.196E+{07} & 20 & 0.849
	& 1.717E+{06} & 2.809E+{06}  
	& 4.894E+{06} \\ 
	$^{212}_{84}$Po& 8.954 & 2.944E{--07} & 22 & 0.925
	& 8.679E{--08} & 1.258E{--07}  
	& 2.620E{--07} \\ 
	$^{214}_{84}$Po& 7.834 & 1.635E{--04} & 22 & 0.929
	& 6.251E{--05} & 8.842E{--05}  
	& 1.732E{--04} \\ 
	$^{216}_{84}$Po& 6.906 & 1.440E{--01} & 22 & 0.931
	& 4.855E{--02} & 7.119E{--02}  
	& 1.400E{--01} \\ 
	$^{218}_{84}$Po& 6.115 & 1.859E+{02} &  22 & 0.932     
	& 4.749E+{01} & 7.354E+{01}  
	& 1.362E+{02} \\ 
	$^{210}_{86}$Rn& 6.159 & 9.000E+{03} & 20  & 0.852
	& 3.697E+{03} & 6.195E+{03}  
	& 9.568E+{03} \\ 
	$^{212}_{86}$Rn& 6.385 & 1.434E+{03} & 20  & 0.845
	& 3.842E+{02} & 6.231E+{02}  
	& 1.049E+{03} \\ 
	$^{214}_{86}$Rn& 9.208 & 2.590E{--07} & 22 & 0.925
	& 1.180E{--07} & 1.755E{--07}  
	& 3.547E{--07} \\ 
	$^{216}_{86}$Rn& 8.198 & 2.900E{--05} & 22 & 0.929
	& 3.543E{--05} & 5.398E{--05}  
	& 1.024E{--04} \\ 
	$^{218}_{86}$Rn& 7.263 & 3.375E{--02} & 22 & 0.931    
	& 2.395E{--02} & 3.704E{--02}  
	& 6.958E{--02} \\ 
	$^{220}_{86}$Rn& 6.405 & 5.560E+{01} &  22 & 0.933    
	& 2.843E+{01} & 4.285E+{01}  
	& 7.681E+{01} \\ 
	$^{222}_{86}$Rn& 5.590 & 3.303E+{05} &  22 & 0.934    
	& 1.232E+{05} & 1.860E+{05}  
	& 3.354E+{05} \\ 
	$^{214}_{88}$Ra& 7.273 & 2.437E+{00} & 20  & 0.841
	& 1.077E+{00} & 1.789E+{00}  
	& 2.883E+{00} \\ 
	$^{216}_{88}$Ra& 9.526 & 1.720E{--07} & 22 & 0.925
	& 1.208E{--07} & 1.761E{--07}  
	& 3.419E{--07} \\ 
	$^{218}_{88}$Ra& 8.540 & 2.991E{--05} & 22 & 0.929    
	& 2.524E{--05} & 3.752E{--05}  
	& 6.835E{--05} \\ 
	$^{218}_{90}$Th& 9.849 & 1.220E{--07} & 22 & 0.925    
	& 1.250E{--07} & 1.742E{--07}  
	& 3.215E{--07} \\ 
	$^{220}_{90}$Th& 8.973 & 1.020E{--05} & 22 & 0.927
	& 1.153E{--05} & 1.725E{--05}  
	& 2.992E{--05} \\ 
	$^{222}_{90}$Th& 8.132 & 2.240E{--03} & 22 & 0.930    
	& 1.911E{--03} & 3.003E{--03}  
	& 5.256E{--03} \\ 
	$^{218}_{92}$U & 8.775 & 3.540E{--04} & 22 & 0.943    
	& 5.625E{--04} & 8.739E{--04}  
	& 1.374E{--03} \\ 
	$^{222}_{92}$U & 9.480 & 4.700E{--06} & 22 & 0.926    
	& 3.517E{--06} & 5.499E{--06}  
	& 9.022E{--06} \\ 
	$^{224}_{92}$U & 8.628 & 3.960E{--04} & 22 & 0.928    
	& 4.763E{--04} & 7.322E{--04}  
	& 1.213E{--03} \\ 
	$^{230}_{94}$Pu& 7.178 & 1.050E+{02} & 22  & 0.937     
	& 1.158E+{02} & 1.844E+{02}  
	& 2.719E+{02} \\ 
	$^{236}_{94}$Pu& 5.867 & 9.019E+{07} & 22  & 0.934     
	& 5.023E+{07} & 8.157E+{07}  
	& 1.089E+{08} \\ 
	$^{238}_{94}$Pu& 5.593 & 2.768E+{09} & 22  & 0.932     
	& 1.344E+{09} & 2.167E+{09}  
	& 2.925E+{09} \\ 
	$^{240}_{94}$Pu& 5.256 & 2.070E+{11} & 22  & 0.931     
	& 1.083E+{11} & 1.791E+{11}  
	& 2.295E+{11} \\ 
	$^{242}_{94}$Pu& 4.984 & 1.183E+{13} & 22  & 0.929     
	& 5.581E+{12} & 8.683E+{12}  
	& 1.125E+{13} \\ 
	$^{244}_{94}$Pu& 4.665 & 2.569E+{15} & 22  & 0.927     
	& 8.072E+{14} & 1.302E+{15}  
	& 1.603E+{15} \\ 
	$^{240}_{96}$Cm& 6.398 & 2.627E+{06} & 22  & 0.929     
	& 1.245E+{06} & 2.005E+{06}  
	& 2.561E+{06} \\ 
	$^{242}_{96}$Cm& 6.216 & 1.407E+{07} & 22  & 0.926    
	& 8.198E+{06} & 1.327E+{07}  
	& 1.714E+{07} \\ 
	$^{244}_{96}$Cm& 5.902 & 5.715E+{08} & 22  & 0.925     
	& 2.737E+{08} & 4.548E+{08}  
	& 5.567E+{08} \\ 
	$^{246}_{96}$Cm& 5.475 & 1.485E+{11} & 22  & 0.924     
	& 5.731E+{10} & 9.442E+{10}  
	& 1.168E+{11} \\ 
	$^{248}_{96}$Cm& 5.162 & 1.199E+{13} & 22  & 0.923     
	& 4.213E+{12} & 7.086E+{12}  
	& 8.335E+{12} \\  
	$^{240}_{98}$Cf& 7.711 & 4.091E+{01} & 22  & 0.925     
	& 6.029E+{01} & 9.635E+{01}  
	& 1.227E+{02} \\ 
	$^{246}_{98}$Cf& 6.862 & 1.285E+{05} & 22  & 0.920     
	& 8.718E+{04} & 1.418E+{05}  
	& 1.747E+{05} \\ 
	$^{248}_{98}$Cf& 6.361 & 2.881E+{07} & 22  & 0.920     
	& 1.448E+{07} & 2.260E+{07}  
	& 2.645E+{07} \\ 
	$^{250}_{98}$Cf& 6.128 & 4.131E+{08} & 22  & 0.918     
	& 1.726E+{08} & 2.856E+{08}  
	& 3.372E+{08} \\ 
	$^{252}_{98}$Cf& 6.217 & 8.614E+{07} & 22  & 0.914     
	& 5.892E+{07} & 9.579E+{07}  
	& 1.079E+{08} \\  
	$^{256}_{104}$Rf& 8.926 & 2.129E+{00} & 22 & 0.905     
	& 1.407E+{00} & 2.344E+{00}  
	& 2.637E+{00} \\ 
	$^{258}_{104}$Rf& 9.196 & 2.551E{--01} & 22 & 0.900     
	& 2.178E{--01} & 3.656E{--01}  
	& 3.928E{--01} \\ 
	$^{260}_{106}$Sg& 9.900 & 1.707E{--02} & 22  & 0.897     
	& 1.493E{--02} & 2.543E{--02}  
	& 2.735E{--02} \\ 
	$^{266}_{108}$Hs& 10.346& 3.947E{--03} & 22  & 0.889     
	& 5.512E{--03} & 8.799E{--03}   
	& 9.091E{--03}  \\ 
	$^{270}_{108}$Hs& 9.070 & 9.000E+{00} & 22  & 0.892     
	& 1.310E+{01} & 2.194E+{01}  
	& 2.170E+{01} \\ 
	$^{270}_{110}$Ds& 11.117& 2.050E{--04} & 22  & 0.883     
	& 3.786E{--04} & 6.363E{--04}  
	& 6.291E{--04} \\  \bottomrule
\end{longtable}

\bibliographystyle{model1a-num-names}

\end{document}